\documentclass[
  aps,
  prd,
  reprint,
  author-year,
  superscriptaddress,
  nofootinbib,
  showkeys,
  showpacs,
  floatfix
]{revtex4-2}

\pdfoutput=1

\usepackage{amsmath,amssymb}
\usepackage{graphicx}
\usepackage{booktabs}
\usepackage{array}
\usepackage{multirow}
\usepackage{dcolumn}
\usepackage{xspace}
\usepackage{siunitx}
\usepackage[american]{babel}
\usepackage[T1]{fontenc}
\usepackage[utf8]{inputenc}
\usepackage{url}
\usepackage[dvipsnames]{xcolor}
\usepackage{float}
\usepackage{tabularx}
\usepackage{ragged2e}
\usepackage[]{hyperref}
\usepackage{placeins}

\hypersetup{
  citecolor = red,
  linkcolor = blue,
  urlcolor = magenta
}

\newcommand{\rd}{r_{\mathrm d}}
\newcommand{\mnu}{\Sigma m_\nu}

\newcommand{\As}{A_s}
\newcommand{\lcdm}{$\Lambda$CDM\xspace}
\definecolor{blueHG}{rgb}{0.0, 0.2, 0.6}
\definecolor{greenW}{rgb}{0.0, 0.55, 0.1}

\newcommand{\tabledashline}[1]{%
  \noalign{\vskip 2pt}%
  \multispan{#1}\leaders\hbox{%
    \vrule width 3pt height 0.4pt\kern 3pt%
  }\hfill\cr
  \noalign{\vskip 2pt}%
}

\begin{document}

\title{Sound-Horizon-Agnostic Forecast for Constraints on $H_0$ and $\Sigma m_\nu$ from BAO, CMB Lensing, Supernovae and LSST-like Galaxy Lensing}

\author{Seyed Hamidreza Mirpoorian}
\email{smirpoor@sfu.ca}
\affiliation{Department of Physics, Simon Fraser University, Burnaby, British Columbia, Canada V5A 1S6}
\affiliation{Laboratoire de Univers et Particules de Montpellier, UMR5299-CNRS, Universite de Montpellier, 34095 Montpellier, France}

\author{Helena Garc\'ia Escudero}
\email{hgarciae@usc.edu}
\affiliation{Department of Physics and Astronomy, University of Southern California, Los Angeles, California, USA}

\author{Levon Pogosian}
\email{levon\_pogosian@sfu.ca}
\affiliation{Department of Physics, Simon Fraser University, Burnaby, British Columbia, Canada V5A 1S6}

\author{Amy Wayland}
\email{amy.wayland@physics.ox.ac.uk}
\affiliation{Department of Physics, University of Oxford, Oxford OX1 3RH, United Kingdom}

\author{David Alonso}
\email{david.alonso@physics.ox.ac.uk}
\affiliation{Department of Physics, University of Oxford, Oxford OX1 3RH, United Kingdom}


\begin{abstract}

We present a sound-horizon-agnostic forecast for constraints on the Hubble constant $H_0$ and the sum of the neutrino masses $\Sigma m_\nu$ from a combination of LSST-like galaxy weak lensing, Simons-Observatory-like CMB lensing, the completed DESI BAO program, and expanded supernova samples. Treating the sound horizon at baryon drag $r_d$ as a free parameter, the method provides a consistency test of $\Lambda$CDM without relying on the standard early-Universe calibration of $r_d$. The analysis incorporates realistic galaxy weak-lensing systematics, explicitly marginalizing over photometric-redshift calibration, multiplicative shear biases, intrinsic alignments, and baryonic feedback, and extends the CMB-lensing analysis to $L_{\max}=5000$. For fixed $\Sigma m_\nu$, we forecast uncertainties of $\sigma(H_0)\simeq 1.5$ km/s/Mpc with our least informative prior set, $\sigma(H_0)\simeq 1.3$ km/s/Mpc with more restrictive priors, and $\sigma(H_0)\simeq 0.7$ km/s/Mpc with an additional informative prior on the amplitude of primordial fluctuations $A_s$. Allowing $\Sigma m_\nu$ to vary weakens the $H_0$ constraints, while still yielding a forecast 95\% CL upper limit of about 0.4 eV on $\Sigma m_\nu$. We also clarify the scope of the $r_d$-agnostic approach through validation tests using mock data from modified recombination and early dark energy models. The method recovers unbiased constraints on $H_0$ for the former but not the latter, highlighting its dependence on assumptions about early-Universe physics.

\end{abstract}

\maketitle

\begingroup
\endgroup

\section{Introduction}
\label{sec:intro}

The determination of the Hubble constant $H_0$ continues to play a central role in assessing the internal consistency of the standard $\Lambda$ Cold Dark Matter (\lcdm) cosmological model. Planck observations of the cosmic microwave background (CMB) analyzed within the context of \lcdm\ yield $H_0=67.36\pm0.54\,\mathrm{km\,s^{-1}\,Mpc^{-1}}$~\cite{Planck:2018vyg}, with other CMB data providing similar results~\cite{AtacamaCosmologyTelescope:2025blo,SPT-3G:2025bzu}, whereas distance-ladder measurements using Cepheid-calibrated supernovae Type IA (SN) infer $H_0=73.17 \pm 0.86\,\mathrm{km\,s^{-1}\,Mpc^{-1}}$~\cite{Breuval:2024lsv}. The resulting discrepancy, now at a $6\sigma$ level, has motivated extensive investigation into both new physics and possible unaccounted-for systematics. In this context, late-time cosmological probes provide a useful avenue for disentangling the assumptions underlying different $H_0$ determinations, particularly those associated with the calibration of the sound horizon at recombination.

Baryon acoustic oscillation (BAO) measurements are among the most robust late-time observables, but their interpretation requires calibration of the sound horizon at the baryon drag epoch, $r_d$, which depends on early-Universe physics. A sound-horizon-agnostic approach avoids this calibration by treating $r_d$ as a free parameter and inferring it jointly with other cosmological parameters. BAO observables constrain quantities such as $D_M(z)/r_d$ and $D_H(z)/r_d$, effectively constraining $H_0 r_d$ and $\Omega_m$, but not $H_0$ independently. Additional probes, such as galaxy and CMB lensing, help break the $H_0$--$r_d$ degeneracy through their sensitivity to growth and clustering, constraining parameter combinations involving $\Omega_m h^2$, $\Sigma m_\nu$ and $\As$, while supernovae provide additional constraints on $\Omega_m$. This approach has been explored in recombination-independent analyses~\cite{Pogosian:2020ded,Pogosian:2024ykm,Sharma:2025iux,Lu:2026uph,Zaborowski:2025umc,GarciaEscudero:2025lef}. In particular, it was shown in \cite{GarciaEscudero:2025lef} that combining the DESI DR2 BAO data with CMB lensing from Planck, ACT and SPT provides competitive constraints on $H_0$ and can also constrain the sum of the neutrino masses $\Sigma m_\nu$.

This paper extends the study in \cite{GarciaEscudero:2025lef}, presenting a forecast of the extent to which near-future galaxy and CMB lensing data, combined with BAO, can improve the $r_d$-agnostic constraints on $H_0$ and $\Sigma m_\nu$. It includes two principal upgrades to the simple forecast performed in \cite{GarciaEscudero:2025lef}. First, we adopt a realistic LSST-like galaxy lensing forecast likelihood based on the methodology of \cite{Wayland:2025udj}, including calibration effects from photometric-redshift uncertainty and multiplicative shear bias. Second, we extend the Simons-Observatory-like (SO-like) CMB-lensing likelihood to higher multipoles, up to $L_{\max}=5000$, using arithmetic binning and updated SO noise assumptions~\cite{Ade:2018sbj}. We combine these with simulated observations corresponding to the completed DESI BAO program~\cite{DESI:2024mwx, DESI:2025zgx} and a SN magnitudes compilation representative of Vera Rubin LSST and Euclid~\cite{Scolnic:2021qvo,Brout:2022vxf}.

We also clarify the scope of interpretation. Treating $r_d$ as a free parameter removes the explicit assumption of a standard early-Universe calibration, but the resulting framework is not fully model-independent. Extensions of $\Lambda$CDM can introduce additional cosmological components and new parameter correlations involving $r_d$ that are not captured by the approach adopted here. We therefore interpret our forecasts within the standard $\Lambda$CDM model, supplemented by a free $r_d$, and discuss this limitation explicitly.

The paper is organized as follows. In Sec.~\ref{sec:rd_agnostic}, we define what we mean by $r_d$-agnostic inference, summarize related approaches in the literature, and state the interpretation caveats. Sec.~\ref{sec:methods} presents the forecast setup, mock-data likelihoods, parameter space, and priors. In Sec.~\ref{sec:validation}, we test the validity of the $r_d$-agnostic method using mock data generated from modified recombination and early dark energy models. Results of our forecasts are presented in Sec.~\ref{sec:results}, followed by the summary and conclusions in Sec.~\ref{sec:conclusions}.

\section{Meaning and Scope of \texorpdfstring{$r_d$}{rd}-agnostic inference}
\label{sec:rd_agnostic}

\subsection{Approaches to determining $H_0$}
{Measurements of the Hubble constant differ not only in statistical precision, but also in the physical assumptions used to evaluate $H_0$ from the observables. Since many proposed early-Universe solutions of the Hubble tension act by changing the sound horizon, it is useful to distinguish measurements that assume a standard sound-horizon calibration from those that avoid using $r_d$ as an external standard ruler. We therefore classify current approaches into three broad categories.

\paragraph{Sound-horizon-calibrated constraints.}

In the standard CMB analysis~\citep{Planck:2018vyg}, $H_0$ is inferred within a specified cosmological model by fitting the full temperature, polarization, and lensing spectra. While the angular acoustic scale plays the central role, the result is model-dependent because it relies on the assumed energy content and recombination physics that determine $r_d$. BAO surveys constrain dimensionless ratios such as $D_M/r_d$ and $D_H/r_d$, which, in combination with an external prediction for $r_d$, yield $H_0$ at the sub-percent level. The precision of these constraints therefore relies explicitly on early-Universe calibration and on the assumption that the cosmological model used to determine $r_d$ also applies to the late Universe. This includes studies in which $H_0$ is inferred from BAO combined with a BBN prior on $\Omega_bh^2$~\cite{Addison:2017fdm}, since they compute $r_d$ using the standard treatment. Any modification to the model used to compute $r_d$, such as early dark energy (EDE)~\cite{Poulin:2018cxd, Poulin:2023lkg}, extra relativistic species~\cite{Vagnozzi:2021gjh}, or modified recombination~\cite{Jedamzik:2020krr,Lee:2022gzh,Lynch:2024gmp,Lynch:2024hzh,Mirpoorian:2024fka,Jedamzik:2025cax,Lee:2026hlo,Lynch:2026pix}, shifts the inferred $H_0$.

\paragraph{Constraints from localized astrophysical sources.}

A second class consists of constraints on $H_0$ derived from localized astrophysical sources without relying on the sound horizon. The most prominent example is the distance ladder, in which geometric anchors calibrate Cepheids, which in turn calibrate Type Ia supernovae in the Hubble flow~\cite{Riess:2021jrx, Breuval:2024lsv}. Alternative implementations of the distance ladder replace Cepheids with the tip of the red-giant branch (TRGB) or J-Region Asymptotic Giant Branch (JAGB) as the primary distance calibrator~\cite{Freedman:2024eph, Freedman:2019jwv}. Complementary methods in this class include one-step geometric approaches such as time-delay strong lensing~\cite{H0LiCOW:2019pvv, Birrer:2020tax} and gravitational-wave standard sirens~\cite{LIGOScientific:2017adf, LIGOScientific:2021aug}, which infer $H_0$ directly from physical observables without relying on a distance-ladder calibration. These methods are independent of $r_d$ and therefore provide an essential comparison to early-Universe-calibrated inferences. However, they are not assumption-free. Their interpretation depends on calibration systematics, population modeling, selection effects, peculiar velocity corrections, and, in some cases, the assumed low-redshift expansion history used to map observed redshifts to distances.

\paragraph{$r_d$-agnostic inference from cosmological survey statistics.}

A third class uses cosmological survey statistics to infer $H_0$ without relying on the standard absolute calibration of the sound horizon. Several implementations have appeared in the recent literature.

One implementation uses full-shape analyses of large-scale structure, in which dependence on $r_s$ is removed by rescaling or marginalizing the oscillatory part of $P(k)$, so that constraints are driven primarily by the smooth component and by the matter-radiation equality scale $k_{\rm eq}$~\citep{Philcox:2020xbv,Philcox:2022sgj,Zaborowski:2024car,Zaborowski:2025umc,Lu:2026uph}.

The method adopted in this work and in~\citep{Pogosian:2020ded,GarciaEscudero:2025lef} uses \emph{uncalibrated} compressed BAO observables which constrain $r_dH_0$ and $\Omega_m$ together with probes sensitive to $\Omega_m h^2$, chiefly CMB lensing and galaxy lensing/clustering, while treating $r_d$ itself as a free sampled parameter. In this approach, the information on $H_0$ is obtained primarily by combining the BAO/SN sensitivity to $\Omega_m$ with lensing sensitivity to $\Omega_m h^2$, allowing $h$ to be inferred without imposing a standard early-Universe calibration of $r_d$. The CMB acoustic angle $\theta_\star$ may be included as an additional high-redshift transverse anchor~\citep{Lin:2021sfs, Pogosian:2024ykm,Zaborowski:2025umc,Sharma:2025iux}. Uncalibrated SN distances can provide a further constraint on $\Omega_m$, though in our current dataset combination their contribution is subdominant once BAO are included.

An approach that is conceptually distinct from the methods described above and is robust to pre-recombination modifications such as EDE was recently proposed in~\cite{Krolewski:2025deb}. The method yields a $\sim3.5\%$ measurement of $H_0$ from the combination of the photon density, the BBN inference of the baryon-to-photon ratio, the baryon-to-matter ratio from clustering, and low-redshift measurements of $\Omega_m$.

The method used in this paper provides the most precise $r_d$-agnostic measurement to date~\cite{GarciaEscudero:2025lef}.

\subsection{Meaning and scope of \boldmath{$r_d$}-agnostic inference}
\label{subsec:rd_agnostic_meaning}

In this paper, ``$r_d$-agnostic'' means that we do \emph{not} adopt the standard \lcdm\ prediction for the drag-epoch sound horizon or calibrate BAO distances to a recombination-derived value of $r_d$. Instead, $r_d$ is treated as a free parameter and inferred jointly with the other cosmological parameters of interest.

This is a statement about \emph{sound-horizon calibration}, not about model independence. Although $r_d$ is left free, the analysis assumes a cosmological model for computing transfer functions, the matter power spectrum, and the growth of structure. Nevertheless, the method provides a valid test of the internal consistency of \lcdm\ across different cosmological observables, regardless of whether the inferred $H_0$ is unbiased in alternative cosmological models. Two caveats follow from this distinction.

First, the method remains sensitive to the assumed early-time cosmology. The use of $\theta_\star$, together with CMB lensing and galaxy-lensing information, assumes that the same cosmological model, without additional early-time components such as EDE or extra relativistic species, consistently describes both the physics around the recombination epoch and the subsequent evolution governing the matter and lensing spectra~\citep{Zaborowski:2025umc,Lu:2026uph,Sharma:2025iux}. For example, CMB lensing probes the broadband shape and amplitude of the matter power spectrum, not only the matter-radiation equality scale $k_{\rm eq}$. Models such as EDE can modify these observables in addition to changing $r_d$. Constraints on $\Sigma m_\nu$ are particularly sensitive in this regard because geometric and growth effects are coupled~\citep{Weiner:2026sfm}.

The extent of the above limitation depends on the class of early-time models being considered. In the modified-recombination scenarios~\cite{Pogosian:2020ded,Lee:2022gzh,Lynch:2024gmp,Lynch:2024hzh,Mirpoorian:2024fka,Jedamzik:2025cax,Lee:2026hlo,Lynch:2026pix}, models that reduce the Hubble tension have the sum of the CDM and baryon densities, $\omega_{cb}$, relatively close to its \lcdm\ value, and the change in the sound-horizon-related parameter combinations is driven primarily by the shift in $r_d$. Treating $r_d$ as a free parameter, therefore, allows one to capture the leading-order parameter dependencies in this class of models. In contrast, EDE models with higher values of $H_0$ induce correlated changes in both $r_d$ and $\omega_{cb}$ that approximately preserve the combination $\omega_{cb}r_d^2$ relevant to the CMB-BAO consistency~\citep{Weiner:2026sfm}. Independently varying $r_d$ within \lcdm does not, therefore, reproduce the parameter dependencies in EDE or, more generally, in models with substantially modified early-time expansion history or the matter perturbation transfer function.

Second, the inferred constraints depend on the adopted priors. In particular, constraints on $\Sigma m_\nu$ remain sensitive to the choice of a uniform versus logarithmic prior when the data do not tightly constrain the parameter. In addition, an informative prior on $A_s$ substantially improves the constraints on both $H_0$ and $\Sigma m_\nu$. We therefore present results obtained with and without an informative $A_s$ prior, and with both a uniform and a logarithmic prior on $\Sigma m_\nu$.

\section{Methods}
\label{sec:methods}

The combination of CMB lensing, galaxy weak lensing, BAO, and SN data allows us to constrain $H_0$ without calibrating $r_d$. The tomographic nature of galaxy weak lensing and the high-$L$ tail of the CMB lensing spectrum help reduce degeneracies involving $A_s$, $\Omega_m h^2$, and $\Sigma m_\nu$. As discussed in Sec.~\ref{subsec:rd_agnostic_meaning}, we assess the dependence of the inferred constraints on the adopted priors, particularly that on $A_s$.

Our methodology involves generating mock data likelihoods based on a fiducial model and anticipated survey specifications, and performing MCMC simulations to determine the forecasted parameter posteriors. We use the Boltzmann code {\tt CAMB}~\cite{Lewis:1999bs} to generate the mock observables and the MCMC sampler Cobaya~\cite{Torrado:2020dgo} for parameter inference. As in \cite{Pogosian:2020ded,Pogosian:2024ykm,GarciaEscudero:2025lef}, we modified {\tt CAMB} and {\tt Cobaya} to treat $r_d$ as a primary (rather than derived) parameter in the analysis of the BAO observables~\footnote{Specifically, in {\tt CAMB}, we introduce a new primary parameter $r_d$ and modify the BAO likelihood file of {\tt Cobaya} ({\tt bao.py}) to use it instead of the derived sound horizon parameter {\tt rdrag}.}. Computations involving observables other than BAO are not modified.

We adopt the Planck best-fit $\Lambda$CDM model as our fiducial cosmology with parameters provided in Table~\ref{tab:fullpriors}. 

\subsection{Mock likelihoods}
\label{subsec:datasets}

\subsubsection{BAO, $\theta_\star$ and SN likelihoods}

To model future BAO, CMB acoustic scale and SN Ia datasets, we adopt the mock data methodology described in~\cite{GarciaEscudero:2025lef}. The BAO data vector is generated for three types of observables: $D_M/r_d$, $D_H/r_d$ and $D_V/r_d$, at 7 effective redshifts used in the DESI DR2 BAO likelihood~\cite{DESI:2025zgx}. We use the fiducial model to compute the central values and adopt the DESI DR2 covariance rescaled by a factor of two, approximating the statistical improvement expected from the completed DESI program over its full 14,000 deg$^2$ footprint~\cite{DESI:2024mwx}.

We also include a measurement of the CMB acoustic scale, $\theta_\star$, defined as the angular size of the sound horizon at the epoch of last scattering~\cite{Pogosian:2024ykm}, $\theta_\star \equiv r_\star / D_M(z_\star)$, where $r_\star$ is the comoving sound horizon evaluated at the redshift of the CMB visibility function peak, $z_\star = 1090$, and $D_M(z_\star)$ is the comoving distance to that redshift\footnote{The comoving distance saturates quickly for $z>100$, so knowing the approximate value of $z_\star$ is sufficient for determining $\theta_\star$.}. Following the approach of~\cite{Pogosian:2024ykm}, we incorporate $\theta_\star$ as an effective transverse BAO-like measurement, $\beta_\star^{\perp} = 94.286 \pm 0.217$ at $z_\star = 1090$, derived from the Planck determination of $\theta_\star$ under the assumption that the standard $\Lambda$CDM relation between $r_d$ and $r_\star$ holds ($r_\star \approx r_d/1.02$); this relation has been verified to remain valid in well-known extensions of $\Lambda$CDM that modify $r_d$.

For the SN forecast, we use the Pantheon+ sample~\cite{Scolnic:2021amr, Brout:2022vxf} as our template and compute the theoretical distance modulus $\mu_{th}(z)$ at each redshift from the fiducial $\Lambda$CDM model. We then generate the mock SN data by adding Gaussian noise assigning uncertainties equal to the Pantheon+ errors reduced by a factor of $2.5$, representing the statistical improvement expected from the future Vera Rubin LSST and Euclid SN~Ia samples~\cite{LSSTDarkEnergyScience:2018jkl, Astier:2014swa}. The resulting mock SN data are compressed into 40 equal-occupancy redshift bins, and the covariance matrix is estimated from 796 independent mock realizations, with the inverse covariance matrix corrected using the Hartlap factor~\cite{Hartlap:2006kj}.

\subsubsection{Simons-Observatory-like CMB lensing}

For the CMB lensing forecast, we follow the mock likelihood construction described in~\cite{GarciaEscudero:2025lef}, with an extended multipole range expected for a Simons-Observatory-like measurement. We construct a Gaussian likelihood for the binned CMB lensing convergence spectrum, $\hat{C_{L_b}^{\kappa\kappa}}$, given by
\begin{equation}
    - 2 \ln \mathcal{L} = \Sigma_{bb'} \left[ \hat{C}_{L_b}^{\kappa\kappa} - C_{L_b}^{\kappa\kappa}(\theta) \right] \mathbb{C}_{bb'}^{-1} \left[ \hat{C}_{L_b'}^{\kappa\kappa} - C_{L_b'}^{\kappa\kappa}(\theta) \right],
    \label{eq:cmb_like}
\end{equation}
where $b$ labels the multipole bin centered at $L_b$, $C_{L_b}^{\kappa\kappa}(\theta)$ is the theoretical binned convergence spectrum for cosmological parameters $\theta$, $\hat{C}_{L_b}^{\kappa\kappa}$ is the mock measured spectrum, and $\mathbb{C}_{bb'}$ is the generated covariance matrix.

Unlike~\cite{GarciaEscudero:2025lef}, where the binning and multipole range were chosen to match the ACT DR6 lensing likelihood covering up to $L_{\rm max} = 2100$, here we extend the lensing multipole coverage to $40 < L < 5000$. We divide the multipole range into $N_{\rm bins} = 50$ non-overlapping bins whose widths follow an arithmetic progression, producing narrower bins at low multipoles and progressively wider bins at high multipoles. This choice captures the increased high-$L$ reach expected from future CMB lensing measurements while avoiding overly fine binning in the noise-dominated small-scale regime.

The covariance matrix is constructed assuming Gaussian uncertainties around the fiducial convergence spectrum, $C_L^{\kappa\kappa}$, with standard deviation given by
\begin{equation}
    \Delta C^{\kappa\kappa}_L = \sqrt{\frac{2}{f_{\rm sky} (2L + 1)}} \left( C^{\kappa\kappa}_L + N^{\kappa\kappa}_L\right),
    \label{eq:Delta_cl}
\end{equation}
where $N_{L}^{\kappa\kappa}$ is the noise power spectrum, and $f_{\rm sky}$ is the fraction of sky coverage~\cite{Knox:1995dq}. For the noise power spectrum $N_{\rm L}^{\kappa\kappa}$, we use the Simons Observatory minimum-variance lensing noise curve {\tt v3.1.1} with the goal sensitivity for the large aperture telescope (LAT)~\cite{SimonsObservatory:2018koc}. Following the methodology described in~\cite{ACT:2023dou}, we estimate the band-power covariance matrix from $N_{\rm mock} = 796$ independent Gaussian realizations, assuming a Simons-observatory-like sky coverage of $f_{\rm sky} = 0.5$.

To account for the biased estimate of the covariance matrix caused by the finite number of simulations, we rescale the inverse covariance by the Hartlap factor~\cite{Hartlap:2006kj},
\begin{equation}
    f_{\rm H} = \frac{N_{\rm mock} - N_{\rm bins} - 2}{N_{\rm mock} - 1}.
    \label{eq:hartlap}
\end{equation}

\subsubsection{LSST-like weak lensing}

For the weak lensing forecast, we use an LSST-like cosmic shear mock dataset based on the pipeline developed in~\cite{Wayland:2025udj}. The generated mock data correspond to a Stage-IV weak lensing survey with specifications motivated by the 10-year LSST sample. The source galaxy population is divided into five tomographic redshift bins with equal number density, and the data vector consists of auto- and cross-correlation of the cosmic shear E-mode angular power spectra between all redshift bin pairs.

The theoretical cosmic shear spectra are computed under the Limber approximation~\cite{Limber:1954},
\begin{equation} \label{eq:wl_cl}
      C_{\ell}^{ij} = G_\ell^2 \int \frac{\mathrm{d} \chi}{\chi^2} \, q_i(\chi) q_j(\chi) P_{\mathrm{mm}}\left(k=\frac{\ell + 1/2}{\chi}, \, z(\chi) \right),
\end{equation}
where $i$ and $j$ label tomographic source bins, $q_i(\chi)$ is the lensing kernel, $P_{\rm mm}$ is the nonlinear power spectrum, and $G_\ell$ accounts for the geometric prefactor relating the three-dimensional gravitational potential to the angular shear field. The mock data are generated using the Core Cosmology Library ({\tt CCL})~\cite{CCL}.

The mock data vector contains the shear power spectra evaluated over the multipole range used in~\cite{Wayland:2025udj}, up to $\ell_{\rm max} = 2000$. The covariance matrix is constructed using the Gaussian Knox formula,
    \begin{equation}
      \mathrm{Cov}(C_\ell^{ij}, C_{\ell'}^{km}) = \frac{\delta_{\ell\ell'}}{f_{\mathrm{sky}}(2\ell+1)} \left(C_\ell^{ik} C_\ell^{jm} + C_\ell^{im} C_\ell^{jk}\right),
      \label{eq:wl_cov}
    \end{equation}
where $f_{\rm sky} = 0.4$ is the fraction of the sky covered by the weak lensing survey.

Following the methodology described in~\cite{Ruiz-Zapatero:2023tnt}, we numerically implement the analytical marginalization over photometric-redshift uncertainties and multiplicative shear calibration biases by modifying the covariance matrix according to 
\begin{equation}
    \widetilde{\mathbb{C}} = \mathbb{C} + \mathbb{T}\mathbb{P}\mathbb{T}^{T},
    \label{eq:wl_tpt_cov}
\end{equation}
where matrix $\mathbb{T}$ contains the derivatives of the theoretical prediction with respect to the calibratable nuisance parameters, and $\mathbb{P}$ is the covariance matrix of the calibratable systematics which is assumed to be uncorrelated. We assume Gaussian priors with widths $\sigma(\Delta z_i)=0.001(1+\bar{z}_i)$ for the redshift-bin shifts and $\sigma(m_i)=0.01$ for the multiplicative shear biases, consistent with the LSST requirements reported in~\cite{LSSTDarkEnergyScience:2018jkl}. This procedure incorporates the degradation from these calibration uncertainties without explicitly sampling the corresponding nuisance parameters.

To use this mock dataset in our joint forecast analysis, we implement a {\tt Cobaya} likelihood based on the LSST-like pipeline of~\cite{Wayland:2025udj}. The likelihood reads the mock data vector and covariance from an input {\tt SACC} file and evaluates
\begin{equation}
    -2\ln\mathcal{L}_{\rm WL} = \left[\hat{\mathbf{d}}_{\rm WL} - \mathbf{d}_{\rm WL}(\theta) \right]^T
    \widetilde{\mathbb{C}}^{-1}
    \left[\hat{\mathbf{d}}_{\rm WL} - \mathbf{d}_{\rm WL}(\theta) \right],
    \label{eq:wl_like}
\end{equation}
where $\hat{\mathbf{d}}_{\rm WL}$ is the mock LSST-like cosmic shear data vector and $\mathbf{d}_{\rm WL}(\theta)$ is the corresponding theory prediction.

For each point in parameter space, the likelihood constructs a {\tt CCL} cosmology using the input cosmological parameters and computes the nonlinear power spectrum. Baryonic feedback is included by applying the {\tt Baccoemu}~\cite{baccoemu} baryonic boost to the nonlinear matter power spectrum, using the seven baryonic parameters
$(\log_{10} M_{\mathrm{c}}$, $\log_{10} \eta$, $\log_{10} \beta$, $\log_{10} M_1$, $\log_{10} \theta_{\mathrm{inn}}$, $\log_{10} \theta_{\mathrm{out}}$, $\log_{10} M_{\mathrm{inn}})$.
Intrinsic alignments are modeled with a redshift-dependent amplitude,
\begin{equation}
    A_{\rm IA}(z) = A_{\rm IA,0} \left(\frac{1+z}{1+z_\star}\right)^{\eta_{\rm IA}},
    \qquad
    z_\star = 0.62,
    \label{eq:wl_ia_amp}
\end{equation}
and are included directly in the weak lensing tracer kernels. The baryonic feedback and intrinsic alignment parameters are treated as nuisance parameters and marginalized over, with priors specified in Sec.~\ref{sec:params}.

While the theory predictions for CMB lensing, BAO, and SN Ia are computed using the modified CAMB, the galaxy weak-lensing likelihood passes the sampled parameter vector to CCL, which internally computes the cosmic shear spectra including baryonic feedback and intrinsic alignments.

\subsection{Parameter and priors}
\label{sec:params}

\begin{table*}[!htbp]
\centering
\begin{ruledtabular}
\begin{tabular}{lcccc}
Parameter & Fiducial value & $P_1$ & $P_1+A_s$ & $U_3$ \\
\colrule

\multicolumn{5}{c}{\textbf{Cosmological parameters}} \\
\colrule

$\ln(10^{10} A_s)$
& 3.045
& $[2,4]$
& $\mathcal{N}(3.045,0.02)$
& $[2.95,3.15]$ \\

$n_s$
& 0.966
& $\mathcal{N}(0.966,0.02)$
& $\mathcal{N}(0.966,0.02)$
& $[0.90,1.02]$ \\

$100\,\Omega_b h^2$
& 2.24
& $\mathcal{N}(2.24,0.05)$
& $\mathcal{N}(2.24,0.05)$
& $[2.05,2.45]$ \\

$H_0$ [km\,s$^{-1}$\,Mpc$^{-1}$]
& 67.32 & $[40,100]$ & $[40,100]$ & $[40,100]$ \\

$\Omega_c h^2$
& 0.120 & $[0.001,0.99]$ & $[0.001,0.99]$ & $[0.001,0.99]$ \\

$\Sigma m_\nu$ [eV]
& 0.06 & $[0,3.0]$ & $[0,3.0]$ & $[0,3.0]$ \\

$\ln(\Sigma m_\nu/1{\rm eV})$
& $\ln(0.06)$ & $[\ln(10^{-3}),\ln(3)]$ & $[\ln(10^{-3}),\ln(3)]$ & $[\ln(10^{-3}),\ln(3)]$ \\

$r_d$ [Mpc]
& 147.1 & $[100,200]$ & $[100,200]$ & $[100,200]$ \\

$\tau$
& 0.054 & fixed & fixed & fixed \\

\colrule
\multicolumn{5}{c}{\textbf{Baryonic feedback parameters}} \\
\colrule

$\log_{10} M_c$
& 14.0 & $[9.0,15.0]$ & $[9.0,15.0]$ & $[9.0,15.0]$ \\

$\log_{10} \eta$
& -0.3 & $[-0.7,0.7]$ & $[-0.7,0.7]$ & $[-0.7,0.7]$ \\

$\log_{10} \beta$
& -0.22 & $[-1.0,0.7]$ & $[-1.0,0.7]$ & $[-1.0,0.7]$ \\

$\log_{10} M_1$
& 10.674 & $[9.0,13.0]$ & $[9.0,13.0]$ & $[9.0,13.0]$ \\

$\log_{10} M_{\mathrm{inn}}$
& 13.0 & $[9.0,13.5]$ & $[9.0,13.5]$ & $[9.0,13.5]$ \\

$\log_{10} \theta_{\mathrm{inn}}$
& -0.86 & $[-2.0,0.5]$ & $[-2.0,0.5]$ & $[-2.0,0.5]$ \\

$\log_{10} \theta_{\mathrm{out}}$
& 0.25 & $[0.0,0.5]$ & $[0.0,0.5]$ & $[0.0,0.5]$ \\

\colrule
\multicolumn{5}{c}{\textbf{Intrinsic alignment parameters}} \\
\colrule

$A_{\mathrm{IA},0}$
& 0.0 & $\mathcal{N}(0,1.0)$ & $\mathcal{N}(0,1.0)$ & $\mathcal{N}(0,1.0)$ \\

$\eta_{\mathrm{IA}}$
& 0.0 & $[-5,5]$ & $[-5,5]$ & $[-5,5]$ \\

\end{tabular}
\end{ruledtabular}
\caption{
Fiducial values and priors used in the forecast analysis. Uniform priors are denoted by square brackets, while Gaussian priors are written as $\mathcal{N}(\mu, \sigma)$. $P_1$ and $P_1+A_s$ denote prior sets similar to those used in~\cite{GarciaEscudero:2025lef}, with $P_1+A_s$ replacing
the uniform prior on the amplitude of primordial fluctuations with a Gaussian prior. $U_3$ adopts uniform priors on $\ln(10^{10}A_s)$, $n_s$, and $100\,\Omega_b h^2$ over the intervals listed above. For each prior set, we consider both fixed and varying neutrino mass.}
\label{tab:fullpriors}
\end{table*}

In addition to the standard cosmological parameters, \{$H_0$, $\Omega_c h^2$, $\Omega_b h^2$, $n_s$, $\ln(10^{10}A_s)$\}, we vary $r_d$ as an independent parameter, and present analyses with and without varying $\Sigma m_\nu$. We also vary and marginalize over the nuisance parameters describing baryonic feedback and intrinsic alignments relevant for LSST weak lensing, following Ref.~\cite{Wayland:2025udj}. The fiducial parameter values used to generate the mock likelihoods are given in Table~\ref{tab:fullpriors}, along with the priors adopted in the MCMC analyses.

We present results for 3 prior sets, labeled $P_1$, $P_1+A_s$, and $U_3$, as summarized in Table~\ref{tab:fullpriors}. The $P_1$ set adopts a broad uniform prior on the amplitude of primordial fluctuations, $\ln(10^{10} A_s) \in [2, 4]$, together with Gaussian priors on $n_s$ and $\Omega_b h^2$, similar to those used in~\cite{GarciaEscudero:2025lef}. The $P_1+A_s$ set replaces the uniform prior on $A_s$ with an informative Gaussian prior, $\ln(10^{10}A_s) \in \mathcal{N}(3.045, 0.02)$. This Gaussian prior is centered on the Planck $\Lambda$CDM value with a width is chosen to encompass the $A_s$ posteriors obtained in a broad range of modified recombination models~\cite{Mirpoorian:2024fka}. A similar Gaussian prior has also been used in another recent $r_d$-agnostic analysis~\cite{Sharma:2025iux}, and helps to break the degeneracy between $A_s$, $\Omega_mh^2$, and $\mnu$, as discussed in~\cite{GarciaEscudero:2025lef}. To further assess the sensitivity of our results to the assumed priors, we consider the $U_3$ set, which adopts uniform priors $\ln(10^{10}A_s)\in[2.95,3.15]$, $n_s\in[0.90,1.02]$, and $100,\Omega_bh^2\in[2.05,2.45]$. The priors on all other parameters are common to these 3 sets.

\section{Validation of the $r_d$-agnostic method.}
\label{sec:validation}

\begin{figure*}[htbp!]
    \centering
    \includegraphics[width=0.7\textwidth]{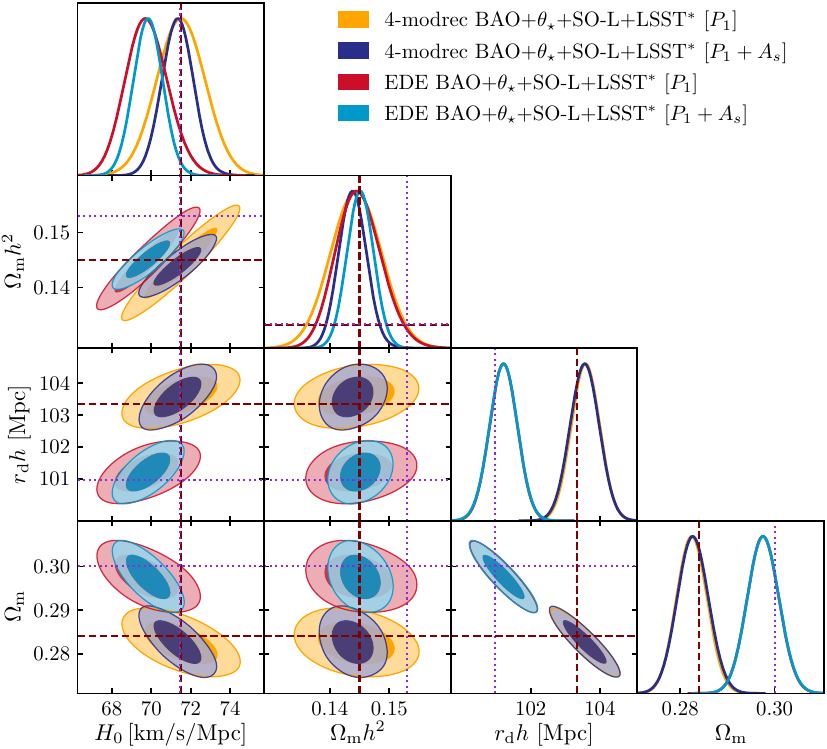}
    \caption{The forecasted 68\% and 95\% CL posteriors for $H_0$, $\Omega_m h^2$, $r_d h$, and $\Omega_m$ obtained from mock SO-like CMB lensing, BAO, $\theta_*$, and $\rm LSST^*$(LSST with fixed nuisance parameters) generated from the modified recombination ({\tt 4-modrec}) and EDE models. The maroon dashed and the blue-violet dotted lines represent the parameter values of the fiducial {\tt 4-moderec} and EDE models. The comparison demonstrates the range of applicability of $r_d$-agnostic method and the impact of the informative $A_s$ prior.}
    \label{fig:triangle_4mod_ede}
\end{figure*}

\begin{table*}[htbp!]
    \centering
    \begin{tabular}{l|ccccccc}
        \hline
        \hline
        {\bf Model} (fixed $\Sigma m_\nu$)
        & $H_0\,(^*)$
        & $100\,\Omega_m h^2$
        & $100\,\Omega_b h^2$
        & $\Omega_m$
        & $r_d h$ [Mpc]
        & $\ln(10^{10} A_s)$
        & $n_s$
        \\
        \hline
        {\tt 4-modrec} fiducial model
        & 71.49
        & 14.54
        & 2.267
        & 0.284
        & 103.35
        & 3.029
        & 0.965
        \\
        {\tt 4-modrec} $\rm BTSL^*$ [$P_1$]
        & $71.5 \pm 1.2$
        & $14.44 \pm 0.40$
        & $2.243 \pm 0.049$
        & $0.283 \pm 0.003$
        & $103.58 \pm 0.39$
        & $3.038 \pm 0.038$
        & $0.964 \pm 0.011$
        \\
        {\tt 4-modrec} $\rm BTSL^*$ [$P_1+A_s$]
        & $71.3 \pm 0.76$
        & $14.38 \pm 0.22$
        & $2.243 \pm 0.051$
        & $0.283 \pm 0.003$
        & $103.56 \pm 0.40$
        & $3.043 \pm 0.018$
        & $0.966 \pm 0.006$
        \\
        {\tt 4-modrec} $\rm BTSL^*$ [$U_3$]
        & $71.5 \pm 1.3$
        & $14.47 \pm 0.48$
        & $>2.20$
        & $0.283 \pm 0.003$
        & $103.55 \pm 0.40$
        & $3.040^{+0.038}_{-0.045}$
        & $0.966^{+0.010}_{-0.012}$
        \\
        \tabledashline{8}
        EDE fiducial model
        & 71.45
        & 15.31
        & 2.264
        & 0.300
        & 100.96
        & 3.071
        & 0.985
        \\
        EDE $\rm BTSL^*$ [$P_1$]
        & $69.8^{+0.92}_{-1.1}$
        & $14.50^{+0.31}_{-0.41}$
        & $2.237 \pm 0.050$
        & $0.298 \pm 0.003$
        & $101.21 \pm 0.38$
        & $3.050^{+0.039}_{-0.031}$
        & $0.994^{+0.011}_{-0.007}$
        \\
        EDE $\rm BTSL^*$ [$P_1+A_s$]
        & $69.8 \pm 0.70$
        & $14.51 \pm 0.21$
        & $2.238 \pm 0.050$
        & $0.298 \pm 0.003$
        & $101.21 \pm 0.38$
        & $3.047 \pm 0.018$
        & $0.994 \pm 0.006$
        \\
        EDE $\rm BTSL^*$ [$U_3$]
        & $69.3 \pm 1.1$
        & $14.33 \pm 0.40$
        & $-$
        & $0.298 \pm 0.003$
        & $101.18 \pm 0.38$
        & $3.067^{+0.034}_{-0.025}$
        & $0.999^{+0.010}_{-0.003}$
        \\
        \hline
    \end{tabular}
    \caption{Cosmological parameters recovered using our method, with fixed $\mnu$, from the mock datasets based on the modified recombination ({\tt 4-modrec}) and EDE models. The ``$\rm BTSL^*$'' dataset denotes the BAO+$\theta_\star$+SO-L+$\rm LSST^*$ data combination used for the validation tests of the \texttt{4-modrec} and EDE models, with $\rm LSST^*$ denoting LSST with fixed nuisance parameters. $^*$($H_0$ is provided in units of ${\rm km}\,{\rm s}^{-1}\,{\rm Mpc}^{-1}$).}
    \label{tab:valid_all}
\end{table*}

To examine the scope of applicability of our method, we tested it on mock data generated using two distinct extensions of the standard cosmological model that give acceptable fits to the cosmological data while predicting a higher value of $H_0$. One was the 4-parameter model of recombination (\texttt{4-modrec}})\footnote{\url{https://github.com/hamidmirpoorian/4modrec}} introduced in~\cite{Mirpoorian:2024fka}, with parameters $A_b=0.3$, $z_b=935$, $\sigma_b=155$ and $\Delta z_{\rm shift} = -39.3$, where $\Delta z_{\rm shift}$ represents an overall shift of the unmodified ionization history in redshift, while the other parameters describe a Gaussian-shaped bump with amplitude $A_b$ and width $\sigma_b$ centered at a reference redshift $z_b$. This {\tt 4-modrec} model provides a good fit to the combination of Planck PR3 low-$\ell$ temperature and Polarization spectra~\cite{Planck:2018vyg, Planck:2019nip}, the NPIPE {\tt CamSpec} high-$\ell$ {\tt TTTEEE} likelihood~\cite{Efstathiou:2019mdh, Rosenberg:2022sdy}, PR4 CMB lensing~\cite{Carron:2022eyg}, DESI DR1 BAO likelihood~\cite{DESI:2025fxa}, the Pantheon+ SN dataset~\cite{Scolnic:2021amr, Brout:2022vxf} calibrated using the SH0ES intrinsic magnitude prior of $M_b = -19.253 \pm 0.027$~\cite{Riess:2021jrx}, and has $H_0=71.49$~km/s/Mpc, $\Omega_mh^2=0.145$ and $r_d = 144.6$~Mpc (see Tables I and VI of~\cite{Mirpoorian:2024fka} for more details).
The second is the axion-like EDE model of~\cite{Smith:2019ihp}, with best-fit EDE energy density fraction $f_{\rm EDE}(z_c)=0.11$, peak redshift of the fractional EDE density $z_c=3364$, initial field value $\theta_i=2.73$ and power index of the axion-like potential $n = 3$. This EDE model fits the combined CMB temperature and polarization spectra from Planck~\cite{Planck:2018vyg}, ACT DR6~\cite{AtacamaCosmologyTelescope:2025nti, AtacamaCosmologyTelescope:2025vnj}, and SPT-3G D1~\cite{SPT-3G:2025bzu} (SPA), DESI DR2 BAO~\cite{DESI:2025zgx}, and SH0ES calibrated Dark Energy Survey 5-year sample of SN dataset (DES-Dovekie)~\cite{DES:2025sig}, and has $H_0=71.45$~km/s/Mpc, $\Omega_mh^2=0.153$ and $r_d=141.3$~Mpc.  Notably, the two models are chosen to provide a similar large value of $H_0 \sim 71.5$ km/s/Mpc, while having different best-fit values of $\As$, with $\ln(10^{10}A_s)= 3.03$ for {\tt 4-modrec} and $\ln(10^{10}A_s)= 3.07$ for EDE.

For each of the models above, we generated mock SO-like CMB lensing, LSST-like galaxy lensing, BAO and $\theta_*$ data following the prescriptions described in the previous section, and applied our $r_d$-agnostic MCMC pipeline to check if there is a bias in the recovered cosmological parameters. To save computing time, for the purpose of this test, we fixed the LSST nuisance parameters at their fiducial values.

The main results of these validation tests are summarized in Table~\ref{tab:valid_all} and Fig.~\ref{fig:triangle_4mod_ede}. We find no bias in the recovered parameters for the {\tt 4-modrec}-based mock data for any of the three priors: $P_1$, $P_1+A_s$ and $U_3$. In contrast, we recover substantially lower $H_0$ and $\Omega_m h^2$, and a higher $r_d$, relative to the input EDE model for all three prior sets.

These tests confirm that the $r_d$-agnostic analysis is unbiased for modified recombination models, since they leave the standard mapping between early-time physics and late-time matter clustering and lensing observables unchanged. In contrast, EDE modifies not only the sound horizon but also the broadband shape of the matter and lensing spectra, including the turnover scale, leading to larger shifts and requiring more caution in interpretation. 

Notably, the recovered values of $\Omega_m$ and $\rd h$ are in excellent agreement with the fiducial values in both {\tt 4-modrec} and EDE. This confirms that the bias in the recovered parameters for EDE is caused by the \lcdm model neglecting the impact of the additional energy component on the shape of the CMB and galaxy lensing spectra.

\section{Results}
\label{sec:results}

\begin{figure*}[ht!]
    \centering
    \includegraphics[width=0.7\textwidth]{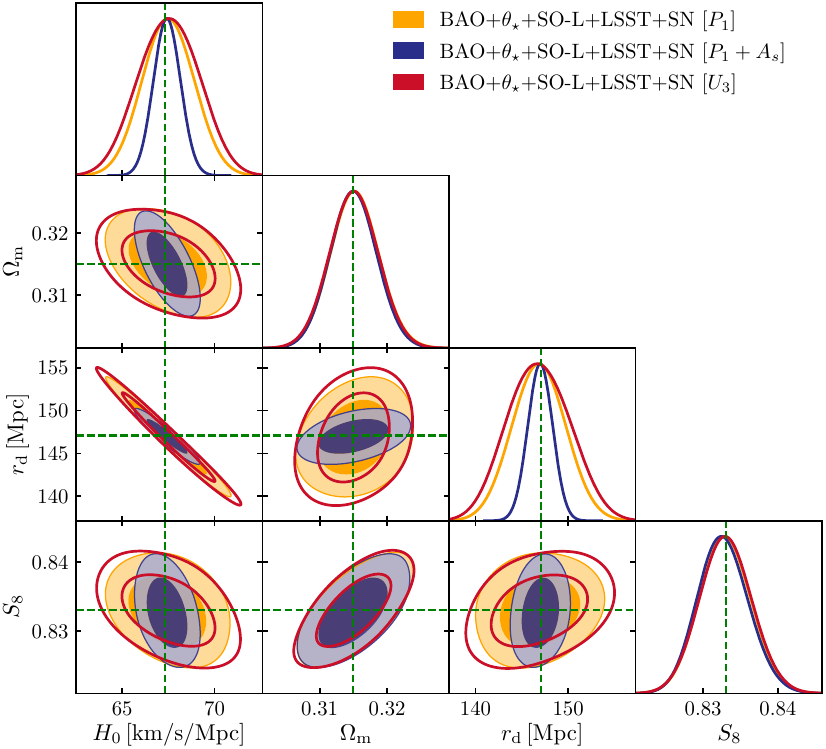}
    \caption{The forecasted 68\% and 95\% CL posterior distributions for $H_0$, $\Omega_m$, $r_d$, and $S_8$ assuming fixed $\mnu = 0.06$ eV. For reference, the green dashed lines represent the fiducial values used in the forecast.}
    \label{fig:triangle_future_fixmnu}
\end{figure*}

\begin{figure*}[ht!]
    \centering
    \includegraphics[width=0.7\textwidth]{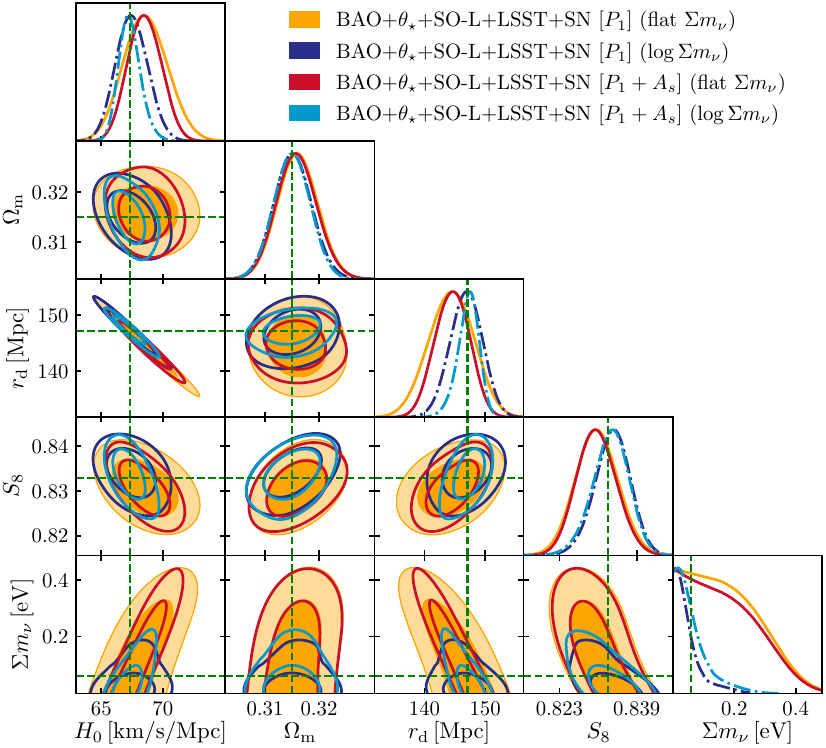}
    \caption{The forecasted 68\% and 95\% CL posterior distributions for $H_0$, $\Omega_m$, $r_d$, $S_8$, and $\Sigma m_\nu$, comparing the results obtained with a uniform and a logarithmic prior on $\mnu$. For reference, the green dashed lines represent the fiducial values used in the forecast.}
    \label{fig:triangle_future_varymnu}
\end{figure*}

\begin{table*}[htbp!]
    \centering
    \begin{tabular}{l|cccccc}
        \hline
        \hline
        {\bf Fixed $\Sigma m_\nu$}
        & $H_0$ [km/s/Mpc]
        & $100\,\Omega_m h^2$
        & $\Sigma m_\nu$ [eV]
        & $r_d$ [Mpc]
        & $\ln(10^{10} A_s)$
        & $S_8$ \\
        \hline
        BTSLS [$P_1$]
        & $67.5 \pm 1.3$
        & $14.36^{+0.48}_{-0.56}$
        & $-$
        & $146.8 \pm 2.7$
        & $3.042 \pm 0.050$
        & $0.833 \pm 0.003$
        \\
        BTSLS [$P_1 + A_s$]
        & $67.4 \pm 0.69$
        & $14.32 \pm 0.23$
        & $-$
        & $147.0 \pm 1.3$
        & $3.045 \pm 0.019$
        & $0.833 \pm 0.003$
        \\
        BTSLS [$U_3$]
        & $67.5 \pm 1.5$
        & $14.37 \pm 0.61$
        & $-$
        & $146.8^{+3.1}_{-3.6}$
        & $3.042^{+0.047}_{-0.072}$
        & $0.833 \pm 0.003$
        \\
        \hline
        \hline
        {\bf Varying $\Sigma m_\nu$} &&&&&& \\
        \hline
        BTSLS [$P_1$]
        & $68.6^{+1.6}_{-1.8}$
        & $14.88^{+0.64}_{-0.85}$
        & $<0.231~ (<0.363)$
        & $144.4^{+3.8}_{-3.4}$
        & $3.044 \pm 0.051$
        & $0.831 \pm 0.004$
        \\
        BTSLS [$P_1$] ($\log \mnu$)
        & $67.5^{+1.2}_{-1.3}$
        & $14.35^{+0.44}_{-0.53}$
        & $0.031^{+0.004}_{-0.032} ~ (0.031^{+0.16}_{-0.03})$
        & $146.9 \pm 2.5$
        & $3.030^{+0.048}_{-0.043}$
        & $0.834 \pm 0.004$
        \\
        BTSLS [$P_1 + A_s$]
        & $68.5^{+1.2}_{-1.4}$
        & $14.83^{+0.49}_{-0.68}$
        & $< 0.227 ~(< 0.358)$
        & $144.0^{+2.5}_{-3.1}$
        & $3.046 \pm 0.019$
        & $0.831 \pm 0.004$
        \\
        BTSLS [$P_1 + A_s$] ($\log \mnu$)
        & $67.3^{+0.64}_{-0.94}$
        & $14.27^{+0.19}_{-0.35}$
        & $0.042^{+0.004}_{-0.045} (0.042^{+0.15}_{-0.048})$
        & $147.2^{+1.9}_{-1.2}$
        & $3.041 \pm 0.017$
        & $0.834^{+0.004}_{-0.003}$
        \\
        BTSLS [$U_3$]
        & $68.7 \pm 1.8$
        & $14.92^{+0.76}_{-0.88}$
        & $< 0.222 ~(< 0.358)$
        & $144.2 \pm 3.8$
        & $3.039^{+0.042}_{-0.072}$
        & $0.830 \pm 0.004$
        \\
        BTSLS [$U_3$] ($\log \mnu$)
        & $67.6 \pm 1.6$
        & $14.40 \pm 0.66$
        & $0.056^{+0.001}_{-0.060} (0.056^{+0.20}_{-0.061})$
        & $146.7 \pm 3.4$
        & $3.039^{+0.043}_{-0.070}$
        & $0.833 \pm 0.004$
        \\
        \hline
    \end{tabular}
    \caption{Forecasted constraints on cosmological parameters with fixed $\mnu = 0.06$ eV, and with varying $\mnu$. For $\mnu$, both 68\% and 95\% CL upper bounds are shown. The ``BTSLS'' dataset denotes BAO+$\theta_\star$+SO-L+LSST+SN, and ``($\log \mnu$)'' denotes the results obtained using a logarithmic prior on $\mnu$ as shown in Figure~\ref{fig:triangle_future_varymnu}.}
    \label{tab:params_all}
\end{table*}

We present the forecasted parameter constraints from the joint analysis of mock BAO, $\theta_\star$, Simons Observatory-like CMB lensing (SO-L), LSST-like galaxy weak lensing, and an expanded SN sample, for three sets of priors: $P_1$, $P_1 + A_s$ and $U_3$. We separately consider the case in which the sum of the neutrino masses is fixed at its minimum value of $\mnu=0.06$~eV and the case in which it is varied as a free parameter. In the latter case, we present results with a uniform and a logarithmic prior on $\mnu$, showing that the prior choice can significantly affect the inferred values of $H_0$, $\Omega_mh^2$, and $\mnu$. The main results are summarized in Table~\ref{tab:params_all} and Figures~\ref{fig:triangle_future_fixmnu} and \ref{fig:triangle_future_varymnu}.

With $\mnu$ fixed, the BAO+$\theta_\star$+SO-L+LSST+SN (BTSLS) dataset with the $P_1$ prior recovers the fiducial value of $H_0$ with an uncertainty of $\sigma({H_0}) = 1.3$~km/s/Mpc. Adding the informative Gaussian prior on $A_s$ reduces it to $\sigma({H_0}) = 0.69$~km/s/Mpc, corresponding to approximately 1\% precision within the class of models consistent with this $A_s$ prior. With the least informative $U_3$ prior, we obtain $\sigma({H_0}) = 1.5$~km/s/Mpc. The corresponding constraints on $\Omega_mh^2$, $r_d$, $A_s$, and $S_8$ are listed in Table~\ref{tab:params_all} and shown in Fig.~\ref{fig:triangle_future_fixmnu}.

Allowing $\mnu$ to vary introduces an additional degeneracy direction and weakens the constraint on $H_0$. The results are sensitive to the choice of the $\mnu$ prior. With $\mnu$ sampled uniformly from $[0,3]$~eV, the BTSLS+$P_1$ analysis gives $H_0 = 68.6^{+1.6}_{-1.8}$~km/s/Mpc and $100\,\Omega_m h^2 = 14.88^{+0.64}_{-0.85}$, with upper limits $\mnu<0.231$~eV at 68\% CL and $\mnu<0.363$~eV at 95\% CL. With the informative Gaussian $A_s$ prior ($P_1+A_s$), we find $H_0 = 68.5^{+1.2}_{-1.4}$~km/s/Mpc and $100\,\Omega_m h^2 = 14.83^{+0.49}_{-0.68}$, with $\mnu<0.227$~eV at 68\% CL and $\mnu<0.358$~eV at 95\% CL. With the $U_3$ prior, we obtain $H_0=68.7\pm 1.8$~km/s/Mpc and $\mnu<0.222$~eV ($<0.358$~eV) at 68\% (95\%) CL. Notably, while still within the $1\sigma$-range of the fiducial model, the recovered mean values of $H_0$ and $\Omega_m h^2$ are consistently larger. This is due to the relatively low constraining power of the BTSLS coupled with the uniform $\mnu$ prior assigning more statistical weight to larger masses. With $H_0$ and $\Omega_mh^2$ being positively correlated with $\mnu$, their marginalized posteriors are consequently displaced toward larger values.

To illustrate the impact of the $\mnu$ prior, we repeat the analysis with $\mnu$ sampled log-uniformly from $[0.001,3]$~eV. For BTSLS+$P_1$, this gives $H_0 = 67.5^{+1.2}_{-1.3}$~km/s/Mpc, $100\,\Omega_mh^2 = 14.35^{+0.44}_{-0.53}$, and $\mnu = 0.031^{+0.004}_{-0.032}$~eV at 68\% CL ($0.031^{+0.16}_{-0.03}$~eV at 95\% CL). Adding the Gaussian prior on $A_s$ yields $H_0 = 67.3^{+0.64}_{-0.94}$~km/s/Mpc, $100\,\Omega_mh^2 = 14.27^{+0.19}_{-0.35}$, and $\mnu = 0.042^{+0.004}_{-0.045}$~eV at 68\% CL ($0.042^{+0.15}_{-0.048}$~eV at 95\% CL). With $U_3$, we obtain $H_0=67.6\pm1.6$~km/s/Mpc and $\mnu=0.056^{+0.001}_{-0.060}$~eV at 68\% CL ($0.056^{+0.20}_{-0.061}$~eV at 95\% CL). While these results are statistically consistent with those obtained using the uniform prior, all parameter posteriors are now closer to their fiducial values.

Our forecast predicts an expected 25--30\% reduction in the $H_0$ uncertainty relative to the current-data results of Ref.~\cite{GarciaEscudero:2025lef}, which applied the same method to DESI DR2 BAO, $\theta_\star$, ACT+Planck+SPT CMB lensing, DES-Year-3 $3\times2$-pt galaxy clustering and lensing, and Pantheon+ SN. For fixed $\mnu$, the current data with the $P_1$ prior yields $H_0 = 70.0 \pm 1.7$~km/s/Mpc, while this forecast predicts an uncertainty of $\sigma(H_0)=1.3$~km/s/Mpc. With the informative $A_s$ prior, Ref.~\cite{GarciaEscudero:2025lef} found $H_0 = 70.03 \pm 0.97$~km/s/Mpc, whereas we forecast $\sigma(H_0)=0.69$~km/s/Mpc.

The gain is more pronounced when $\mnu$ is varied. Using the uniform mass prior for direct comparison, the BTSLS+$P_1$ forecast gives $H_0=68.6^{+1.6}_{-1.8}$~km/s/Mpc and $\mnu<0.363$~eV at 95\% CL, compared with $H_0=75.3^{+3.3}_{-4.0}$~km/s/Mpc and $\mnu<1.1$~eV at 95\% CL from current data. With the informative $A_s$ prior, the forecast gives $H_0=68.5^{+1.2}_{-1.4}$~km/s/Mpc and $\mnu<0.358$~eV at 95\% CL, compared with $H_0=71.5^{+1.4}_{-2.9}$~km/s/Mpc and $\mnu<0.86$~eV at 95\% CL from current data. This improvement in constraining $\mnu$ reflect the added information in the shapes of the LSST-like galaxy lensing and the extended SO-like CMB-lensing spectra.

\section{Conclusions}
\label{sec:conclusions}

In this work, we have extended the sound-horizon-agnostic forecast of~\cite{GarciaEscudero:2025lef} by incorporating a realistic LSST-like galaxy weak-lensing likelihood of~\cite{Wayland:2025udj} that includes nonlinear structure formation and relevant observational and astrophysical systematics. We also extended the SO-like CMB-lensing likelihood to higher multipoles, up to $L_{\rm max} = 5000$.

Our main result is that combining the DESI-like BAO and CMB $\theta_\star$ with the LSST-like galaxy lensing and extended SO-like CMB-lensing will improve the current $r_d$-free constraints~\cite{GarciaEscudero:2025lef} on $H_0$, particularly when $\mnu$ is varied, thanks to the additional information in the shapes of the galaxy and CMB weak-lensing spectra. With $\mnu$ fixed, we forecast an uncertainty of $\sigma(H_0)=1.3$ km/s/Mpc, improving to $0.69$ km/s/Mpc with an informative prior on $A_s$. Allowing $\mnu$ to vary weakens the $H_0$ constraint, but yields a 95\% CL upper limit of approximately $0.36$ eV on the neutrino mass sum. We also find that the inferred values of $H_0$ and $\mnu$ are sensitive to the choice of a uniform versus logarithmic prior on $\mnu$.

We also clarified the sense in which the analysis should be interpreted as ``sound-horizon-agnostic''. The parameter $r_d$ is varied freely, and the inference does not assume its standard value or compute it from theory. However, this does not make the method completely independent of early-Universe physics. In particular, the use of the CMB acoustic angular scale $\theta_\star$ as a high-redshift anchor, together with CMB-lensing spectrum information, implicitly assumes that the mapping between early-time scales and the late-time matter and lensing spectra is correctly described by the assumed baseline model~\cite{Zaborowski:2025umc,Lu:2026uph,Sharma:2025iux}. Our validation tests demonstrate unbiased recovery of $H_0$ for the modified-recombination model considered, but not for the EDE model, which modifies the matter and lensing power spectra in addition to the sound horizon. Importantly, the method remains a valid test of the internal consistency of \lcdm\ across different cosmological observables, regardless of whether it provides unbiased parameter estimates in alternative cosmological models. Its value as a consistency test is therefore more general than its applicability as a parameter inference method.

Overall, our results show that future CMB lensing and next-generation galaxy weak-lensing data will play a central role in sound-horizon-agnostic tests of the Hubble tension. As BAO, CMB lensing, and galaxy weak lensing and clustering measurements become increasingly precise, this $r_d$-agnostic framework will provide an increasingly precise test of whether the Hubble tension can be resolved by modifications to the sound horizon while preserving the standard predictions for late-time matter clustering and lensing.

\section*{Acknowledgments}
This research was enabled in part by support provided by the BC DRI Group and the Digital Research Alliance of Canada ({\tt alliancecan.ca}). SHM and LP are supported by the National Sciences and Engineering Research Council (NSERC) of Canada.   AW is supported
by a Science and Technology Facilities Council (STFC)
studentship. DA acknowledges support from the Beecroft Trust. HGE is supported by the Grant 63667 from the John Templeton Foundation. The opinions expressed in this publication are those of the authors and do not necessarily reflect the views of the John Templeton Foundation. Note that this is not an official Simons Observatory collaboration publication.

\bibliographystyle{apsrev4-2}

%

\end{document}